%% file: main.tex
\documentclass[sigconf]{acmart}

\copyrightyear{2026}
\acmYear{2026}
\setcopyright{cc}
\setcctype{by}
\acmConference[EASE '26]{International Conference on Evaluation and Assessment in Software Engineering}{June 09--12, 2026}{Glasgow, United Kingdom}
\acmBooktitle{International Conference on Evaluation and Assessment in Software Engineering (EASE '26), June 09--12, 2026, Glasgow, United Kingdom}
\acmDOI{10.1145/3816483.3816559}
\acmISBN{979-8-4007-2348-3/2026/06}

\usepackage{graphicx}
\usepackage{booktabs}
\usepackage{multirow}
\usepackage{tcolorbox}

\title{The Replication Assessment Problem in Software Engineering}

\author{Giuseppe Destefanis}
\affiliation{%
  \institution{University College London}
  \country{UK}
}
\email{g.destefanis@ucl.ac.uk}

\author{Martin Shepperd}
\affiliation{%
  \institution{Brunel University of London}
  \country{UK}
}
\email{martin.shepperd@brunel.ac.uk}

\author{Leila Yousefi}
\affiliation{%
  \institution{Ministry of Justice}
  \country{UK}
}
\email{lilyyousefi84@gmail.com}

\begin{abstract}
\textbf{Background:}
Replication studies in software engineering are increasingly common, yet their interpretation remains uncertain and inconsistent because assessments frequently rely on loosely defined or ad hoc criteria.

\textbf{Aim:}
This study aims to document how replication study outcomes are currently assessed in empirical software engineering, identify problems arising from inconsistent criteria and propose a principled framework for meaningful evaluation.

\textbf{Method:}
We conducted a systematic review of replication studies, with the search covering recent empirical software engineering replications (2021--2025). For each study, we extracted the criteria used to assess replication outcomes and analysed these for heterogeneity, logical consistency, and alignment with established statistical principles.

\textbf{Results:}
A total of 10 replication studies were located. The analysis reveals substantial heterogeneity in assessment practices, with contradictory criteria applied to similar data, limited acknowledgement of measurement uncertainty, and an absence of shared standards. We propose a principled framework grounded in statistical, methodological, and measurement considerations, and demonstrate its application through worked examples.

\textbf{Conclusions:}
Adopting consistent and transparent assessment principles would reduce ambiguity, improve comparability, and support more reliable evidence accumulation in software engineering replication research.
\end{abstract}

\begin{document}

\maketitle

\input{introduction}
\input{background}
\input{methodology}
\input{results}
\input{rq3}
\input{discussion}

\input{threats}

\input{conclusion}

\bibliographystyle{ACM-Reference-Format}
\bibliography{biblio}

\end{document}

%% file: introduction.tex
\section{Introduction}
\label{sec:introduction}

Replication is central to the accumulation of reliable empirical knowledge. By subjecting prior findings to independent scrutiny, replication studies allow us to distinguish robust phenomena from statistical artefacts or context-dependent effects. In software engineering, interest in replication has grown substantially, producing taxonomies, reporting guidelines, and systematic investigations of replication practice~\cite{76157cdbd7ca2f0c74eae4aba4ded14985a3e2b6, c361f1d73c6f40e6d1d3abe33f3aa7cd9e0d1b4c, Cruz2020,Gonzalez2023revisiting}. Yet a fundamental problem remains largely unaddressed. How should the outcomes of replication studies be assessed?  How do we decide that the results of a replication are congruent or not?

When a replication reports results that differ from the original, under what conditions should this be interpreted as a failure to replicate, as evidence of boundary conditions or as an artefact of measurement variation? Conversely, when results appear similar, what criteria determine whether the replication has succeeded? Current practice exhibits substantial heterogeneity in how these questions are answered. Researchers employ different metrics, statistical tests, significance thresholds, effect size comparisons, and qualitative criteria when deciding whether a replication confirms or fails to support a prior study~\cite{ShepperdEtAl2018,Heyard2025scoping}. This heterogeneity means that two replication studies with numerically similar results may reach opposite conclusions, depending on the criteria applied.

Shepperd et al.~\cite{ShepperdEtAl2018} provide indirect evidence of this problem. In their review of replications in software effort prediction and pair programming, internal replications reported confirmatory evidence approximately eight times more frequently than external replications. While this disparity may partly reflect legitimate procedural differences, it also raises questions about whether assessment criteria differ systematically and whether such differences are justified. Without agreed criteria for what constitutes a successful replication, claims about the state of replication in software engineering lack an objective basis.

This descriptive study addresses the following research questions:

\begin{description}
    \item[RQ1:] \textit{What criteria are currently used to assess replication outcomes in empirical software engineering, and how do these criteria vary across studies?}
\end{description}

\begin{description}
    \item[RQ2:] \textit{What problems arise from inconsistent or underspecified assessment criteria?}
\end{description}

\begin{description}
    \item[RQ3:] \textit{What principles should guide the assessment of replication outcomes in software engineering research?}
\end{description}

We conducted a search of replication studies published between 2021 and 2025, building on the baseline corpus of Cruz et al.~\cite{Cruz2020}, and identified 10 studies for inclusion. For each study, we extracted and analysed the criteria used to assess replication outcomes. We then go on to propose a set of preliminary assessment principles grounded in statistical and methodological considerations.

Section~\ref{sec:background} reviews related work. Section~\ref{sec:method} describes the method. Section~\ref{sec:results} presents findings on assessment practices and their inconsistencies. Section~\ref{sec:principles} proposes preliminary principles for replication assessment. Section~\ref{sec:threats} discusses threats to validity, while Section 8 concludes the paper.

%% file: background.tex
\section{Background}
\label{sec:background}

A persistent challenge in discussions of replication is terminological inconsistency. Following the ACM Artifact Review and Badging framework, as summarised by González-Barahona and Robles~\cite{Gonzalez2023revisiting}, we use \textit{repetition} for a study with identical setup performed by the same team, \textit{reproduction} for an identical setup performed by a different team, and \textit{replication} for a study with a different setup performed by a different team. We note that Sj{\o}berg et al.~\cite{Sjoberg2005} distinguish \textit{internal replications}, where the replication team includes original authors, from \textit{external replications}, conducted by an entirely independent team. Gómez et al.~\cite{76157cdbd7ca2f0c74eae4aba4ded14985a3e2b6} classify replications as literal, operational, or conceptual, depending on how much of the original protocol is preserved. These distinctions potentially may matter because the appropriate criteria for assessing outcomes may differ across replication types.

Replication studies in software engineering go back to the mid 1990s, e.g., Daly et al.~\cite{Daly1994} in 1994 and Wood et al.,~\cite{Wood1997} in 1997.  Subsequently, there has been growth in the number of replications, although it remains a minority activity (see  \cite{daSilva2014,Cruz2020} for reviews/mapping studies). 

Interestingly, what has received very little explicit attention is how do we determine the outcome of a replication?  Specifically, how close should the results of the replication be to those of the original study in order to constitute `confirmation'? Should there be other outcomes than simply `confirmation' and `disconfirmation'?  If so, what are they and how should we determine our answers?  

In a recent mapping study of empirical research more generally, Heyard et al.~\cite{Heyard2025scoping} examined methodological literature focused on measuring, assessing, explaining, and predicting study reproducibility.  They also included an investigation into what measures were actually used in 49 large-scale replication projects such as \cite{Klein2018many}.  They identified a large number of distinct metrics: 50 in total.  These include classic significance testing, direction of effect, effect size difference, the use of Bayes factors and various subjective assessment frameworks.  The median number of metrics per replication was two, though this ranged 1-12.  It was also noteworthy that almost 30\% of the studies reported using subjective or narrative assessment of reproducibility.  

Furthermore, Nosek and Errington~\cite{nosek2020} argue that given a replication is a study for which any outcome constitutes diagnostic evidence about a prior claim, then the assessment criteria should be specified in \textit{advance} and that both confirmatory and disconfirmatory outcomes should alter confidence in the original finding.   

%Bonett~\cite{Bonett2021} proposed a classification of replication outcomes into types of statistical replication evidence, non-replication evidence, and inconclusive evidence, based on effect compatibility within a specified bandwidth.

In terms of software engineering, we are not aware of much explicit discussion of replication metrics and methods.  Shepperd \cite{Shepperd2018harmful,ShepperdEtAl2018} argued that prediction intervals should be used to evaluate replication outcomes (as opposed to confidence intervals) and showed via simulation the difficulties of meaningfully confirming underpowered  studies.  He, along with Santos et al.~Santos et al., \cite{Santos2021}, also advocated using incremental meta-analysis to pool study results, as initially proposed for psychological studies by Braver et al.~\cite{Braver2014continuously}.

There has been some interest in using Bayesian methods to evaluate replications since this allows us to explicitly model our beliefs concerning prior studies, typically the original study, and also its updating with new data, the replication study(ies) \cite{Ly2019replication}. However, the application in software engineering that we are aware of is a tutorial from Erdogmus~\cite{Erdogmus2022bayesian}.

Fletcher~\cite{Fletcher2021} discusses in detail and critiques different approaches to evaluating replication study results and arguing that there are conceptual or formal reasons weaknesses with all approaches.  A major objection is that by dichotomising the replication result into success or unsuccessful leads to the loss of much useful scientific knowledge.  For this reason he concludes that pooling results for meta-analyses is the least problematic approach, a position which we the authors agree.

None of these frameworks has been systematically applied to assess how replication outcomes are currently judged in software engineering, which is the gap this study addresses.  In the next section we describe our systematic review of replication assessment methods and metrics in software engineering.

% Shepperd et al.~\cite{} used prediction intervals to evaluate whether replication results were compatible with originals, and Shepperd~\cite{a41df9050ef1e4505f70b11b377d60d751e74a9c} argued that single under-powered replications contribute little to knowledge and that meta-analysis is more informative. Santos et al.~\cite{2bba09c650e83ad4986961dc474ffb06f30135ed} showed that direct comparison of p-values and raw effect sizes across replications is unreliable due to sampling error, and recommended treating the baseline experiment as one estimate among many rather than a privileged target. 

%% file: methodology.tex
\section{Method}
\label{sec:method}

\begin{figure*}[htb!]
\centering
\begin{tikzpicture}[
  node distance=2.8cm,
  box/.style={rectangle, draw, text width=2.3cm, align=center, minimum height=1.5cm, font=\small, rounded corners}
]
\node[box, fill=gray!20] (search) {SCOPUS Search\\[0.1cm]2021--2025\\[0.05cm](n = 62)};
\node[box, right of=search, fill=blue!10] (screen) {Screening\\[0.1cm]IC/EC criteria\\Full-text review\\[0.05cm](n = 10)};
\node[box, right of=screen, fill=green!10] (rq1) {RQ1\\[0.1cm]Extract\\assessment\\criteria};
\node[box, right of=rq1, fill=orange!10] (rq2) {RQ2\\[0.1cm]Analyse\\problems};
\node[box, right of=rq2, fill=red!10] (rq3) {RQ3\\[0.1cm]Synthesise\\assessment\\principles};
\draw[-latex, thick] (search) -- (screen);
\draw[-latex, thick] (screen) -- (rq1);
\draw[-latex, thick] (rq1) -- (rq2);
\draw[-latex, thick] (rq2) -- (rq3);
\end{tikzpicture}
\caption{Methodological workflow showing sequential stages from corpus identification to framework synthesis.}
\label{fig:methodology}
\end{figure*}
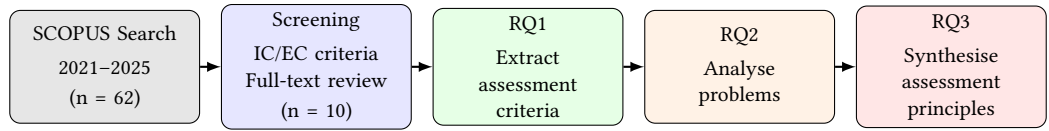

This study employs a systematic approach to identify assessment practices in replication studies (RQ1), analyse problems arising from inconsistent criteria (RQ2), and synthesise principles for assessment (RQ3). Figure~\ref{fig:methodology} gives an overview of the methodological workflow.

\subsection{Study Identification}

\paragraph{Search strategy.}
We searched SCOPUS in February 2026 using the following query, with year constraints \texttt{pubyear > 2020 AND pubyear < 2026}:

\begin{quote}
\small
\texttt{"software engineering" AND title-abs-key("experiment*" OR "case stud*" OR "observational stud*" OR "pilot stud*" OR "survey") AND title-abs-key("repli*" OR "family of*")}
\end{quote}

\noindent
This query follows the strategy used by Cruz et al.~\cite{Cruz2020}, who identified 137 replication studies published between 2013 and 2018 using the same terms, however, we focused on recent practise, i.e., research published in the past five years.  Our search returned 62 candidate papers.

\paragraph{Inclusion and exclusion criteria.}
We included studies that: (IC1) report at least one replication of an empirical study in software engineering; and (IC2) were published between 2021 and 2025. We excluded studies where: (EC1) the term replication refers to another context (e.g., data replication in distributed systems); (EC2) replication is mentioned only as future work; (EC3) the document is not a research paper (e.g., extended abstract, tutorial); or (EC4) the study is not in English. For each candidate paper, we verified that the study presents newly conducted analysis or data collection, and that empirical results are reported. Studies that revisit or comment on previously published work without new data were excluded.

\paragraph{Screening and data extraction.}
One researcher screened titles, abstracts, and full texts of all 62 retrieved papers against the inclusion and exclusion criteria, yielding 10 included studies. For each included study, the researcher extracted: (1) bibliographic details; (2) the outcome type (e.g., continuous, binary, ordinal); (3) the criteria used to assess replication outcomes, coded as 13 binary method indicators; (4) the replication verdict as reported by the authors; and (5) whether the basis for that verdict was clearly stated.

Our 13 replication assessment methods (derived from \cite{Fletcher2021}) are: null hypothesis significance testing for difference; equivalence testing; directional equivalence testing; confidence interval comparison; prediction intervals; effect size comparison; directional or trend assessment only; predictive accuracy comparison; expert or narrative judgement; pooling via meta-analysis; Bayesian methods; other formal methods; and methods too poorly described to classify. These codes were derived deductively from the literature and applied to each included study. Whether the original study appeared in the same paper as the replication was also recorded. The full coding guide and dataset are available in the replication package at this link: \textbf{\href{https://figshare.com/s/dd573fa5465ad21d97fd}{Figshare repository}}.

\subsection{Analysis}

To address RQ1, we report the frequency distribution of the 13 assessment method codes across the 10 included studies, and examine how many methods each study employed. To address RQ2, we identify cases where the basis for a verdict is absent or inconsistent with the methods used, and discuss the consequences for evidence accumulation. To address RQ3, we synthesise findings from RQ1 and RQ2 with the statistical and interpretive literature reviewed in Section~\ref{sec:background}, drawing in particular on Bonett~\cite{Bonett2021} and Nosek and Errington~\cite{nosek2020}, to derive a set of preliminary assessment principles. To characterise heterogeneity in assessment approaches, studies were additionally grouped thematically based on their dominant method.

%% file: results.tex
\section{Results}
\label{sec:results}

\subsection{Dataset}

\begin{sloppypar}
The replication package provides the full coding dataset as a XLXS file, where each row corresponds to a candidate paper from the SCOPUS search. Columns include bibliographic details (e.g., \texttt{paper\_id}, authors, title, year, source, DOI), inclusion decisions (e.g., \texttt{in\_scope\_se}, \texttt{presents\_as\_replication}, \texttt{included\_final}), and, for included studies, outcome characteristics and assessment data. The latter comprise 13 binary method indicators (\texttt{m\_*}), the reported replication verdict, whether its basis was clearly stated, whether the original study appeared in the same paper and reviewer notes. Of the 62 retrieved papers, 52 were excluded and 10 retained for analysis. 
\end{sloppypar}

\subsection{RQ1: Assessment Criteria in Current Practice}

Table~\ref{tab:methods} shows the assessment methods used across the 10 included studies, along with the reported verdict, outcome type, and whether the judgement basis was clearly stated.

% new
\begin{table*}[htb!]
\centering
\caption{Assessment methods and verdicts for included studies grouped by dominant assessment approach. Method columns use Yes/No coding. NHST-D: significance test for difference; NHST-E: equivalence testing; NHST-ES: directional equivalence; ES: effect size comparison; PA: predictive accuracy; EJ: expert judgement; PO: pooling; OF: other formal; UN: unclear.}
\label{tab:methods}
\small
\begin{tabular}{lclcccccccccll}
\toprule
ID & Reference & Outcome & NHST-D & NHST-E & NHST-ES & ES & PA & EJ & PO & OF & UN & Verdict & Basis \\
\midrule

\multicolumn{14}{l}{\textbf{Formal statistical comparison}} \\
S016 & \cite{Duran2024} & Mixed      & -- & -- & Y  & -- & -- & -- & -- & -- & -- & Inconclusive & Yes \\
S036 & \cite{Bernardez2020} & Continuous & Y  & -- & Y  & Y  & -- & -- & Y  & -- & -- & Successful   & Yes \\
S049 & \cite{Langhout2021} & Binary     & -- & Y  & -- & -- & -- & -- & -- & -- & -- & Partial      & Yes \\

\midrule
\multicolumn{14}{l}{\textbf{Expert judgement}} \\
S006 & \cite{Koana2024} & Continuous & -- & -- & -- & Y  & -- & Y  & -- & -- & Y  & Partial      & No \\
S017 & \cite{Barclay2024} & Continuous & -- & -- & -- & -- & -- & Y  & -- & -- & Y  & Successful   & No \\
S033 & \cite{dosSantos2022} & Continuous & -- & -- & -- & -- & -- & Y  & -- & -- & -- & Partial      & Yes \\

\midrule
\multicolumn{14}{l}{\textbf{Pooling}} \\
S011 & \cite{Cornejo2024} & Ordinal    & -- & -- & -- & -- & -- & Y  & Y  & -- & -- & Partial      & Unclear \\
S023 & \cite{Aranda2022} & Continuous & -- & -- & -- & -- & -- & -- & Y  & -- & Y  & None         & No \\
S053 & \cite{Diaz2021} & Continuous & -- & -- & -- & -- & -- & -- & Y  & Y  & -- & Partial      & Yes \\

\midrule
\multicolumn{14}{l}{\textbf{Predictive accuracy}} \\
S026 & \cite{Fan2023} & Continuous & -- & -- & -- & -- & Y  & -- & -- & -- & -- & Failed       & Yes \\

\bottomrule
\end{tabular}
\smallskip
\end{table*}

Expert judgement and pooling were the most frequently used methods, each appearing in four studies. Effect size comparison and directional equivalence testing each appeared in two studies. Null hypothesis significance testing for difference, standard equivalence testing, predictive accuracy comparison, and other formal methods each appeared in one study. Four methods were not used by any included study: confidence interval comparison, prediction intervals, directional assessment only, and Bayesian methods. The number of methods used per study ranged from one (S016, S023, S026, S033, S049) to four (S036). Five studies relied on a single method to reach their verdict.

Seven of the 10 studies involved continuous outcomes. The remaining three involved ordinal, binary, and mixed outcome types respectively. Five studies assessed a replication conducted within the same paper as the original study, and five assessed an externally published replication.

The 10 studies can be grouped into four clusters according to their dominant assessment approach, which directly illustrates the heterogeneity in current practice. The first cluster comprises studies that use at least one formal statistical method for direct comparison: S036 applies null hypothesis significance testing, directional equivalence testing, effect size comparison, and pooling; S049 uses equivalence testing based on odds ratios derived from a side-by-side comparison of binary outcomes across programming languages; and S016 applies directional equivalence testing and explicitly acknowledges power limitations, reaching an inconclusive verdict. These three studies are the most methodologically explicit in the corpus. The second cluster comprises studies where expert or narrative judgement is the dominant or sole method: S006 compares correlation coefficients in detail but does not state a threshold for what difference would constitute failure; S017 reaches a successful verdict on the basis of expert judgement with no stated criteria; and S033 conducts seven external replications and reports a partial verdict through narrative assessment. The third cluster comprises studies where pooling is the primary analytical step: S011 pools data across three replications without a separate analysis of the first study, S023 conducts a baseline experiment followed by 11 internal replications with no explicit replication verdict, and S053 uses replication number as a blocking factor in a mixed model to justify aggregation. In all three cases, the question of whether individual replications are compatible with the original is not directly addressed. The fourth cluster contains a single study, S026, which assesses replication through predictive accuracy comparison, reflecting the tool-oriented nature of the research rather than a hypothesis about an effect.

\subsection{RQ2: Problems Arising from Inconsistent Criteria}

Of the 10 included studies, six reported a clearly stated basis for their verdict, three did not, and one was unclear. This means that in four cases the link between the evidence and the conclusion cannot easily be verified by a reader.

The most common verdict was partial or mixed, reported by five studies. Two studies reported a successful replication, one reported failure, one reported an inconclusive outcome, and one provided no explicit judgement despite involving a baseline experiment followed by 11 internal replications (S023). In S023, the stated goal was to justify pooling all available data rather than to assess whether individual replications confirmed the original finding, illustrating how pooling can displace rather than support replication assessment.

Three specific problems are evident in the data. Expert judgement is the sole assessment method in S033, and the only classifiable method in S017, where the basis for the verdict was not stated in sufficient detail to allow classification of all methods used. In S017, a verdict of successful replication is reached with no stated basis, and the methods used are coded as unclear. This means an independent reader cannot verify the basis for the verdict from the reported information alone.

Second, the absence of stated thresholds renders verdicts unverifiable. S006 compares correlation coefficients across studies but does not state what magnitude of difference would constitute a failure to replicate, rendering the basis for the partial verdict unclear to an independent reader. S036, by contrast, uses four methods including effect size comparison and equivalence testing and provides a clear basis for its successful verdict, making it the most transparent study in the corpus.

Third, pooling is used as a substitute for direct replication comparison in three studies (S011, S023, S053). In each case, the primary analytical goal is to aggregate data across replications rather than to assess whether individual replication results are compatible with the original. Pooling serves a legitimate purpose in families of experiments, but it does not answer the question of whether a given replication succeeded or failed.

No study used prediction intervals, confidence interval comparison, or Bayesian methods to assess replication outcomes, despite these being the approaches most directly suited to expressing effect compatibility under uncertainty~\cite{Bonett2021}.

%% file: rq3.tex
\section{Preliminary Principles for Replication Assessment}
\label{sec:principles}

The findings of RQ1 and RQ2 identify four recurring problems: assessment criteria are frequently unstated, expert judegment is used without explicit decision rules, pooling displaces direct replication comparison, and methods well-suited to expressing effect compatibility are absent from current practice. Drawing on the literature reviewed in Section~\ref{sec:background}, we propose four preliminary principles to address these problems.

\paragraph{Principle 1: State assessment criteria before examining results.}
Nosek and Errington~\cite{nosek2020} argue that declaring a study to be a replication entails a commitment to treating all outcomes as diagnostic evidence. This commitment requires that the criteria for success and failure are specified in advance, not derived post hoc from the observed results. Four of the 10 studies in our corpus either do not state or are unclear about the basis for their verdict. Pre-specification of criteria would make verdicts independently verifiable and reduce the risk that thresholds are selected to favour confirmatory conclusions.

\paragraph{Principle 2: Express verdicts in terms of effect compatibility, not significance alone.}
Two studies (S016, S036) use directional equivalence testing, and one (S049) uses standard equivalence testing based on odds ratios; S036 additionally compares effect sizes. No study uses prediction intervals or confidence interval comparison. Bonett~\cite{Bonett2021} provides a classification of replication outcomes based on whether the replication effect falls within a specified bandwidth of the original, distinguishing statistical replication evidence, non-replication evidence, and inconclusive evidence. Many researchers \cite{Shepperd2018harmful,Fletcher2021} have demonstrated that direct comparison of p-values is unreliable due to sampling error and instead recommended treating the original as one estimate among many. Verdicts grounded in effect compatibility, with explicitly stated bandwidths, are more informative and more reproducible than verdicts based on whether both studies reach the same significance threshold.

\paragraph{Principle 3: Distinguish pooling from replication assessment.}
Three studies (S011, S023, S053) use pooling as the primary analytical step without first assessing whether individual replication results are compatible with the original. Pooling is appropriate once compatibility has been established or as a means of obtaining a more precise aggregate estimate, but it does not itself constitute evidence that a replication succeeded or failed. Studies that report only a pooled result without a direct comparison between the original and replication findings leave the assessment question unanswered.

\paragraph{Principle 4: Report inconclusive outcomes explicitly.}
Only S016 reports an inconclusive verdict, the patterns observed in the corpus are consistent with several others having done so. S023 reaches no explicit judgement despite a substantial family of replications, and S017 reports a successful verdict with methods that could be better described to classify. Bonett~\cite{Bonett2021} identifies inconclusiveness as a legitimate outcome category arising from wide intervals or insufficient power. Reporting inconclusive outcomes explicitly, rather than resolving ambiguity through narrative judgement, would produce a more accurate picture of the state of evidence in a research programme.

\paragraph{Worked example.}
To illustrate how the four principles (P1-P4) apply in practice, consider S017~\cite{Barclay2024}, which investigates consistency of gender bias in machine translation tools over time and reports a successful replication verdict. The methods used are coded as expert judgement and unclear, and the basis for the verdict is not stated.

Under P1, the authors would be required to specify in advance what pattern of results across time points would constitute confirmation and what would constitute failure. Without this, the successful verdict cannot be independently verified from the reported information alone. Under P2, the verdict would need to be grounded in a quantitative comparison of effect magnitudes across time points, with an explicit statement of the bandwidth within which the effects would be considered compatible. Under P3, no pooling is involved, so this principle does not apply. Under P4, given that the methods are insufficiently described to classify and no basis is stated for the verdict, the assessment under the proposed principles would be inconclusive rather than successful. The outcome is not that the replication failed, but that the available reporting does not provide sufficient information to reach a verdict. This distinction matters for evidence accumulation: an inconclusive outcome signals that further evidence is needed, whereas a successful verdict without a stated basis does not provide sufficient information to calibrate confidence in the robustness of the original finding.

%% file: discussion.tex
\section{Discussion}
\label{sec:discussion}

The findings from the 10 included studies are consistent with the concern raised by Shepperd et al.~\cite{ShepperdEtAl2018} about systematic differences in how replication outcomes are judged. In their analysis, internal replications reported confirmatory evidence approximately eight times more frequently than external replications. Our data do not allow us to test this disparity directly, given the small corpus size, but the prevalence of unstated criteria and expert judgement as the sole assessment method in several studies is consistent with the hypothesis that verdicts are influenced by factors other than the evidence itself. When the basis for a verdict is not stated, it is not possible to determine whether a confirmatory conclusion reflects genuine compatibility between original and replication results or reflects a permissive interpretation of ambiguous data.

The dominance of expert judgement and pooling as assessment approaches reflects a broader tendency in SE replication research to treat families of experiments primarily as opportunities for data aggregation rather than for testing the robustness of specific claims. This is not without value: pooling increases statistical power and can yield more precise effect estimates. However, as Santos et al.~\cite{Santos2021} argue, the baseline experiment should be treated as one estimate among many rather than as a privileged target. Aggregation should nonetheless follow from, not substitute for, an assessment of whether individual replications are compatible with prior findings. The three studies in our corpus that use pooling as the primary step (S011, S023, S053) do not first establish this compatibility.

The absence of prediction intervals, confidence interval comparison, and Bayesian methods from all 10 studies is notable given that these are the approaches most directly suited to the problem of assessing effect compatibility under uncertainty. Shepperd et al.~\cite{ShepperdEtAl2018} demonstrated the use of prediction intervals for this purpose, and Bonett~\cite{Bonett2021} provides a formal classification framework built on effect bandwidth. The gap between available methodology and current practice suggests that adoption barriers exist, whether related to familiarity, software availability, or reporting norms, that are worth addressing in future work.

The definition proposed by Nosek and Errington~\cite{nosek2020}, in which a replication is a study for which any outcome constitutes diagnostic evidence about a prior claim, has a direct implication for assessment: if both confirmatory and disconfirmatory outcomes are to be taken seriously, then the criteria for distinguishing them must be stated in advance. Only six of the 10 studies in our corpus meet this condition. The remaining four either assert a verdict without a stated basis or leave the basis unclear, which means their outcomes cannot function as diagnostic evidence in the sense intended by Nosek and Errington. Adopting pre-specified criteria, as proposed in P1, would bring current SE replication practice closer to this standard.

%% file: threats.tex
\section{Threats to Validity}
\label{sec:threats}

\paragraph{Internal validity.}
All 62 candidate papers were screened and coded by a single researcher. This introduces a risk of coding error and subjective interpretation, particularly for fields that require judgement, such as \texttt{judgement\_basis\_clear} and the \texttt{notes} field. A multi-coder design with inter-rater reliability assessment would reduce this risk. The coding guide and dataset are provided in the replication package to allow independent verification.
Coder confidence, recorded during extraction, was rated high for five studies (S006, S016, S033, S036, S053), medium for four (S011, S017, S023, S049), and low for one (S026), suggesting that coding uncertainty is concentrated in a small number of cases.

\paragraph{External validity.}
The corpus consists of 10 studies retrieved from a single database (SCOPUS) covering 2021--2025. Studies indexed only in other databases, or published outside this period, are not represented. The findings are sufficient to identify recurring patterns and illustrate specific problems, but they do not support claims about the prevalence of these problems across SE replication research more broadly. The search query follows Cruz et al.~\cite{Cruz2020}, which provides comparability with prior work, but the query may miss replication studies that do not use the terms \textit{replication} or \textit{family of} in their title, abstract, or keywords.

\paragraph{Construct validity.}
The 13 method codes were derived deductively from the literature and applied to the reported content of each study. In some cases, the methods used by authors were insufficiently described to allow unambiguous classification, as reflected in the \texttt{m\_unclear} code assigned to two studies. Where authors use informal or narrative language to describe their assessment approach, mapping this to a binary code inevitably involves interpretation.

%% file: conclusion.tex
\section{Conclusions}
\label{sec:conclusions}

Replication studies in software engineering are of limited value if their outcomes cannot be compared, aggregated, or interpreted on consistent terms. This descriptive study documented how replication outcomes are currently assessed in 10 empirical software engineering studies and identified four recurring problems: unstated assessment criteria, reliance on expert judgement without decision rules, use of pooling as a substitute for direct replication comparison, and absence of methods suited to expressing effect compatibility under uncertainty. Four preliminary principles were proposed to address these problems, grounded in the statistical and interpretive literature on replication assessment.

The corpus is small and the findings are intended as a basis for further investigation rather than as definitive claims about SE replication practice. Future work should extend the search to a larger corpus, apply a multi-coder design, and test the proposed principles on a broader range of replication studies. The coding guide and dataset are provided in the replication package to support such extensions.